\documentstyle[psfig,namedreferences]{kluwer}
\def\deg{\ifmmode^\circ\else$^\circ$\fi}

\runningtitle{ EVOLUTION  OF   MAGNETIC   NONPOTENTIALITY }
\runningauthor{YONG-JAE MOON ET AL.}
\begin{opening}
\title{EVOLUTION OF MAGNETIC NONPOTENTIALITY ASSOCIATED WITH
AN X-CLASS FLARE IN AR 6919}
\author{Y.-J. \surname{MOON}\thanks{e-mail:yjmoon@boao.re.kr}}
\institute{Korea Astronomy Observatory, Whaamdong, Yooseong-ku, Taejon, 
305-348, Korea}
\author{H. S. \surname{YUN}}
\institute{Department of Astronomy, Seoul National University, Seoul 151-742,
Korea}
\author{G. S. \surname{CHOE}}
\institute{ Princeton Plasma Physics Laboratory, Princeton, NJ 08543-0451, USA}
\author{Y. D. \surname{PARK}}
\institute{Korea Astronomy Observatory, Whaamdong, Yooseong-ku, Taejon, 305-348, Korea}
\author{D. L. \surname{MICKEY}}
\institute{Institute for Astronomy, University of Hawaii, 2680
Woodlawn Drive, Honolulu, HI 96822-1839, USA}
\date{}
\end{opening}
\begin{document}

\begin{abstract}
We present the evolution of magnetic 
nonpotentiality associated with an 
X-class flare in AR 6919 using a set of MSO (Mees Solar Observatory) magnetograms. 
The magnetogram data were obtained before and after the flare, 
using the Haleakala Stokes Polarimeter which provides  
simultaneous Stokes profiles of the 
Fe I doublet 6301.5 and 6302.5. A nonlinear least square 
method was adopted to derive the magnetic field configuration from the 
observed Stokes profiles and a multi-step ambiguity solution 
method was employed to resolve the 
$180\deg$ ambiguity. 

From the $180\deg$ ambiguity-resolved vector magnetograms, 
we have derived a set of physical quantities characterizing  
the field configuration such as  vertical current 
density, magnetic shear angle, angular shear and magnetic free energy 
density.  We have examined their changes before and after the flare 
occurrence and obtained the following results. 1) There was  
a  flux decrease  
in both polarities around  a $\delta$-type sunspot region, where an X-class flare occurred.
2) The vertical  current near the  $\delta$ sunspot region
 was  strongly enhanced before  the 
flare. 3) The magnetic shear  near the $\delta$ sunspot region increased  before the flare and 
then decreased after it. 
4) The sum of magnetic energy density significantly 
decreased before the flare onset, implying that
magnetic free energy was released through the 
flaring processes. 
\end{abstract}

\section{Introduction}
It is generally believed that the energy released in solar
flares is stored in  nonpotential magnetic fields. 
The energy buildup could be a response of the coronal magnetic field to 
the changes in photospheric magnetic environment caused mostly by 
sunspot motions and emerging fluxes.
Since the measurement of magnetic fields in the corona 
is not available, the field measurement at the photospheric level
has been widely used
for the study of magnetic nonpotentiality in
flare-producing active regions.
In particular, the change of some nonpotentiality parameters
during flare activity is regarded as a very important clue 
for understanding flare mechanisms.

The MSFC (Marshal Space Flight Center) group studied
the magnetic nonpotentiality associated
with solar flares using  the MSFC magnetograph 
(Hagyard {\it et al.}, 1982; Hagyard {\it et al.}, 1984).
They examined the vertical current density, source field and 
magnetic shear derived from vector 
magnetograms of active regions in parallel with
flare observations
(Hagyard {\it et al.}, 1984;
Gary {\it et al.}, 1987; Hagyard {\it et al.}, 1990).
On the basis of these studies,
they  suggested a few important points characterizing
flare producing active regions. They are
a large shear angle, a strong transverse magnetic field along
the neutral line and a twist of long neutral lines
(Gary {\it et al.}, 1991). The
BBSO (Big Bear Solar Observatory) and Huairou (Beijing Observatory)
groups have also presented
several interesting studies on the
change of nonpotentiality parameters associated with
solar flares (e.g., Wang {\it et al.}, 1996;  Wang 1997).
Wang (1992) defined a transverse field 
weighted angular shear and made a first attempt
to study  the change of the weighted mean
shear angle just after a flare from vector magnetograph
measurements. He found that the weighted mean shear angle
jumped about 5 degrees, coinciding
with a flare.
He also showed that five  more X-class and M-class
flares had the same pattern of shear angle variation 
(see Table  I in
Wang 1997).  It is to be noted that the shear increases
in some active regions  during flare activities
coincide with emerging
flux loops (e.g., Wang and Tang, 1993). 
On the other hand, the MSFC group
reported that the shear
may increase, decrease or remain the same
after major flares (Hagyard, West, and Smith, 1993;
Ambastha, Hagyard, and West, 1993).
Also, Chen {\it et al.} (1994)
observed that there were no detectable changes in magnetic shear
after 18 M-class flares. 
Thus, the time variation of magnetic shear before and
after a solar flare still remains controversial.

For a quantitative understanding of nonpotentiality
parameters such as angular shear, 
we have to keep in mind that there are several
technical problems calling for a careful treatment. 
First, the calibration from polarization signals to vector fields
should be well established. 
Second, the $180\deg$ ambiguity of the observed transverse field
has to be properly resolved. 
Most of nonpotentiality parameters such as vertical current density
and magnetic shear critically depend on how well
the $180\deg$ ambiguity problem is solved.
Third,
it is noted that Stokes line profiles could be changed due to
the variation of thermodynamic parameters
associated with flaring processes.

So far, studies on the change of magnetic shear have been
made 
by using mainly filter-based magnetographs thanks to their wide field
of views and high time resolutions (Zirin, 1995; Wang {\it et al.}, 1996; Wang, 1997).
However, filter-based magnetographs have some problems
which originate from insufficient spectral information. 
Wang {\it et al.} (1996)
well reviewed reliability and limitations of filter-based magnetographs.
The calibration of filter-based
magnetographs has often been made by  employing the line slope method (Varsik,  1995) 
under the weak field approximation
(Jefferies and Mickey, 1991) or  using nonlinear
calibration curves based on  theoretical atmospheric models
(Hagyard {\it et al.}, 1988; Sakurai {\it et al.}, 1995;  Kim, 1997).
In these methods,
the longitudinal and transverse field components
are independently given by 
\begin{equation}
B_L = K_L {V \over I} \, ,
\end{equation}
\begin{equation}
B_T = K_T \biggl [ {Q^2+U^2 \over I^2} \biggr ]^{1/4} \, ,
\end{equation}
in which $K_L$ and $K_T$ are calibration coefficients or curves
which depend on the observing system or the models considered.
Hagyard and Kineke (1995) developed an iterative method for  Fe I 5250.22 line 
by considering theoretical polarization signals with various inclination angles of the 
magnetic field.
Recently, Moon, Park, and Yun (1999b) have devised an iterative calibration method
for Fe I 6302.5 line following  Hagyard and Kineke (1995). 
They applied this calibration method  to 
 the dipole model of Skumanich (1992)
in determining  
the field configuration of a simple round sunspot.
This study revealed that the conventional calibration methods remarkably
underestimate transverse field strengths in the inner penumbra.

Most of previous studies on magnetic shear have adopted
the potential method for resolving the $180\deg$ ambiguity (Wang {\it et al.}, 1994).
However, the method may possibly break down for flaring active regions,
which generally have a strong shear near the polarity inversion line. 
In filter-based magnetographs, magnetic field vectors are
derived from polarization signals integrated over the filter passband and
their calibration relations are determined without considering 
variation of thermodynamic parameters. That is to say,
the derived magnetic fields may be affected by various physical 
processes involved in flares.

In this study, we use a set of MSO magnetograms of AR 6919 obtained by
the Haleakala Stokes  Polarimeter which provides  simultaneous full  Stokes polarization 
profiles. 
Since the calibration
is based on the nonlinear least square method
(Skumanich and Lites, 1987), the  magnetic field
 components derived are  expected to be 
less affected by flare-related physical processes. 
Since it takes about an hour to scan 
a whole active region by the polarimeter, 
we can not detect such an abrupt
change of magnetic shear as Wang (1992) found.
In our study, emphasis is given to  
the evolution of magnetic nonpotentiality with the progress of an X-class flare
which occurred in AR 6919.
A similar study for AR 5747 is being
prepared by Moon {\it et al.} (1999c).
 We expect that 
our data are able to yield a more reasonable estimate 
of various nonpotentiality parameters.
In Section 2, an account  is given of the  magnetic nonpotentiality parameters that  we have 
considered  in  this study. 
The observation and analysis are presented in Section 3  and the evolution of nonpotentiality  
parameters  is discussed in Section 4. 
Finally, a brief summary and conclusion are delivered in Section 5.  

\section{Magnetic Nonpotentiality Parameters}

\subsection{Electric Current Density}

It is well accepted that   electric  currents  play  an     important     role   in the    
process  of  energy buildup and relaxation  
of solar  active   regions. 
Since the  observation   of vector   magnetic 
fields is available 
only in  the    photosphere,     
the  vertical current    at      the photosphere  is  widely  
used in studies of solar active   regions. 
According to Ampere's law, 
the vertical current density is given by
\begin{equation}
J_z = {1 \over \mu_0} \biggl ( {\partial B_y \over \partial x} -
{ \partial B_x \over \partial y} \biggr ) \, ,
\end{equation}
in which 
$\mu_0=0.012 \rm Gm/A$ is the magnetic permeability of free space.

\subsection{Magnetic Angular Shear}
Hagyard {\it et al.} (1984) defined
the magnetic angular shear (or magnetic shear) as the
angular difference   between the  observed  transverse field   and the  azimuth  of the 
transverse component of 
the potential field which is computed employing the observed longitudinal field 
as a boundary condition.  
That is, the magnetic angular shear
$\theta_a$  is given by
\begin{equation}
\theta_a=\theta_o - \theta_p \, ,
\end{equation}
in which $\theta_o= \arctan(B_y/B_x)$ is the azimuth of observed
transverse field and
$\theta_p= \arctan(B_{px}/B_{py})$ is that of the corresponding
potential field component.
Noting that flares are associated with magnetic shear of strong transverse
fields, Wang (1992) proposed a transverse weighted mean angular shear 
given by
\begin{equation}
\overline \theta_a  = {\sum B_t \theta_a \over \sum B_t} \, ,
\end{equation}
in which $B_t$ is the transverse field strength and the sum is taken over all the
pixels in the considered region. In this work  we use horizontal field strength
in the heliographic coordinate
instead of transverse field strength.

\subsection{Magnetic Shear Angle}

L\"u {\it et al.} (1993) suggested a new nonpotentiality indicator,
the angle between the observed magnetic field vector and
the corresponding potential magnetic field vector.
By definition, the shear angle $\theta_s$ can be expressed as
\begin{equation}
\theta_s=\arccos \biggl ( {{\bf B_o} \cdot {\bf B_p} 
\over  |{\bf B_o}| |{\bf B_p}|} \biggr )
\end{equation}
where 
${\bf B_o}$ and
${\bf B_p}$ are 
the observed and  potential magnetic field vectors, respectively.  In this equation, 
$B_{pz}$ is identical with 
$B_{oz}$ as explained earlier. 
In our study, we proceed a  step further and consider a   field strength weighted mean 
shear angle defined by
\begin{equation}
\overline \theta_s  = {\sum |{\bf B}| \theta_s \over \sum |{\bf B}|}
\end{equation}
where $|{\bf B}|$ is the field strength and the sum is taken over all the
pixels in the considered region.

\subsection{Magnetic Free Energy Density}

The density of magnetic free energy is given by 
\begin {equation}
\rho_f =  { ({\bf B_o} - {\bf B_p})^2 \over 8 \pi }
      =  { {\bf B_s}^2 \over 8 \pi }
\end{equation}
where ${\bf B_s}$ is the nonpotential part 
of the magnetic field, which was defined as the 
source field by Hagyard, Low, and Tandberg-Hanssen (1981).
It can also be expressed  as (Wang {\it et al.}, 1996)
\begin{equation}
\rho_f = {(|{\bf B_o}| - |{\bf B_p}|)^2 \over 8 \pi }
      +  {|{\bf B_o}| |{\bf B_p}| \over 2 \pi} \sin ^2 (\theta_s/2) \, ,
\end{equation}
where $|{\bf B_{o}}|$ and $|{\bf B_{o}}|$ are magnitudes of the observed
field 
and the computed potential field, respectively.
The tensor virial theorem can be 
utilized to estimate the total magnetic free energy
of a solar active region (e.g., Metcalf {\it et al.}, 1995). 
However, as McClymont et   al. (1997) pointed out  (for details,  see   Appendix A of  
their paper), 
there
are several controversial problems in estimating the magnetic free
energy of a real active region with 
this method. 
In this study,  we examine an observable  quantity, the sum  of magnetic free 
energy density over a field  of view. This   quantity is expected to indicate the degree 
of nonpotentiality at least near the photosphere.

\section{Observation and Analysis}

For the present work, we have selected 
a set of MSO magnetogram data of  AR 6919
taken on Nov. 15, 1991.
The magnetogram data were obtained by
the Haleakala Stokes polarimeter (Mickey, 1985) 
which provides simultaneous Stokes I,Q,U,V profiles  
of the Fe I 6301.5, 6302.5 ${\rm \AA}$  doublet. 
The observations were made by a rectangular raster scan
with a pixel spacing of 2.8$''$ (high resolution
scan) or 5.6$''$ (low resolution scan)
and a dispersion
of 25 ${\rm m \AA / pixel}$. Most of the analyzing procedure is  
well described in Canfield {\it et al.} (1993).
To derive the magnetic fields from Stokes
profiles, we have used a nonlinear least square fitting 
method (Skumanich and Lites, 1987) for fields stronger than 
100 G and an integral method (Ronan, Mickey, and
Orral 1987)
for weaker fields.
In that fitting, 
the Faraday rotation effect, which is one of the error sources
for strong fields, is properly taken into account. 
The noise level in the original magnetogram is about 70 G for
transverse fields and 10 G for longitudinal fields.
The basic observational parameters of magnetograms used in this study are
presented in Table I.

To resolve the $180\deg$ ambiguity and to transform
the image plane (longitudinal
and transverse)
magnetograms to heliographic (vertical and horizontal)
ones,  
we have adopted a multi-step 
method by Canfield {\it et al.} (1993) (for details, see the
Appendix of their paper). In Steps 3 and 4 of their method,   
they have chosen the orientation of the transverse field 
that minimizes the angle
between neighboring field vectors and the field divergence 
$|\nabla \cdot {\bf B}|$.
Moon {\it et al.}  (1999a) have  already discussed   in detail  the problem  of $180\deg$ 
ambiguity
resolution  for present magnetograms.

\begin{table}
\caption{Basic observational parameters of AR 6919. }
\begin{tabular}{ccccc}\hline 
Data  &  Date &  Time  & 
Scan Resolution &  Data Points  \\
\hline
AR 6919 a) &  Nov. 15, 1991 & 17:50-18:55 & 5.656$''$ & 35$\times$30\\
AR 6919 b) &  Nov. 15, 1991 & 21:05-22:45 & 2.828$''$ & 60$\times$45\\
AR 6919 c) & Nov. 15, 1991 & 23:46-25:25 & 2.828$''$ & 60$\times$45\\
\hline
\end{tabular}
\end{table}

\section{Evolution of Magnetic Nonpotentiality}

Moon {\it et al.} (1999a) already explained in detail three magnetograms
under consideration and their 
characteristics.
Here we present  the second 
image plane vector magnetogram of AR 6919 superposed 
on white light images in Figure 1. 
According to the Solar Geophysical  Data, a 
3B H$\alpha$  flare  and an X1.5 flare occurred around   22:35    UT   on November 
15, 1991 with the
heliographic coordinate of S13 and W19. 
Considering the timing of the flaring events and the weak projection effect,
our selected magnetograms are 
useful for studying the change of magnetic field  structures
associated with the flare.  This active region was also studied in terms of coordinated Mees/Yohkoh observations (Canfield {\it et al.}, 1992; W\"ulser {\it et al.}, 1994),
 X-ray imaging observations (Sakao {\it et al.}, 1992),  white light flare observations (Hudson {\it et al.}, 1992), and its preflare phenomena (Canfield and Reardon, 1998).
From the three vector magnetograms in heliographic
coordinates, we have derived a set of 
nonpotential
parameters described in Section 2.
In this study, we pay attention to  their time variation 
associated
the X-class flare.

\begin{figure}
\psfig{figure=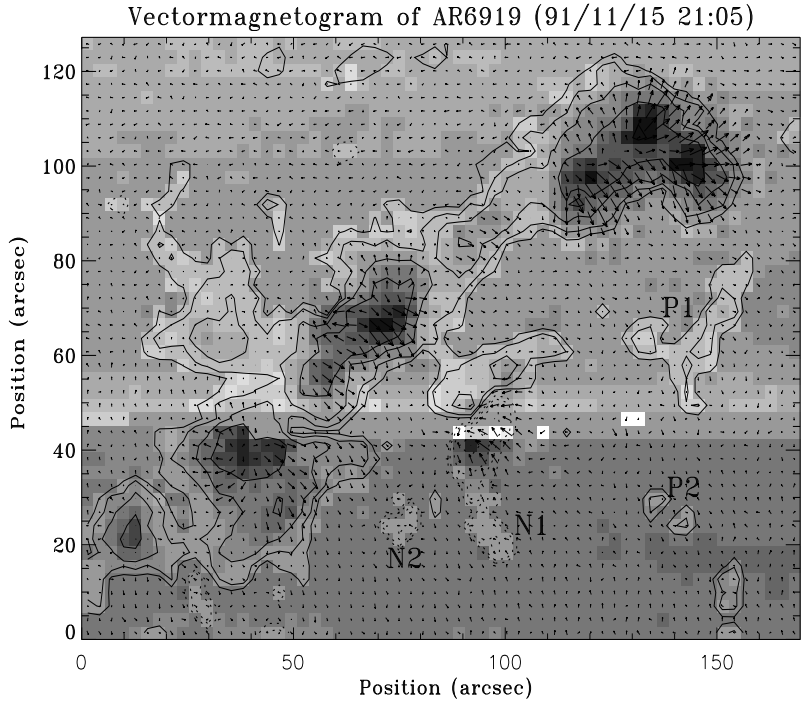,height=10cm,width=12cm}
\caption{
An image plane MSO vector magnetogram of AR 6919  superposed on
white light images taken on November 15, 1991.
In the figure, the  solid lines stand for positive polarities and  the   dotted lines for 
negative polarities. The contour levels correspond to 100, 200, 400, 800 
and 1600 G, respectively.  The   length of arrows    
 represents   the magnitude of   transverse 
field component.}
\end{figure}  

\begin{figure}
\psfig{figure=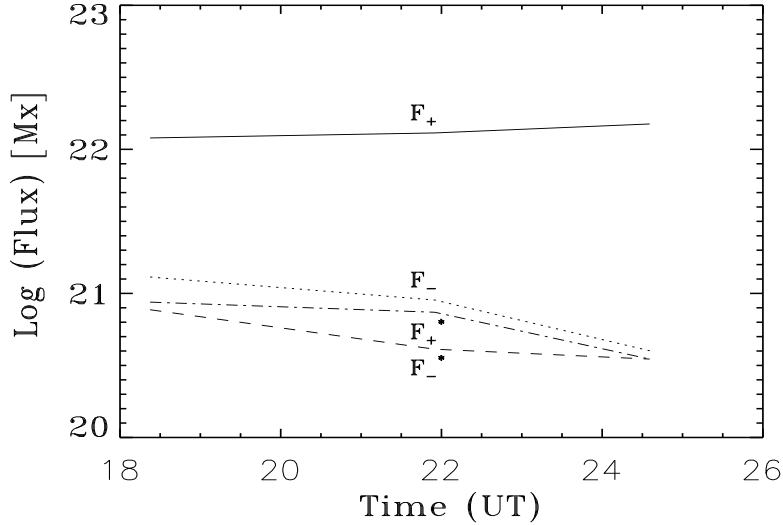,height=8cm,width=12cm}
\caption{Time variation  of  magnetic fluxes   before and   after an  X-class  flare in  
AR 6919. 
The
curves marked with an asterisk ($*$) stand for fluxes in a $\delta$ sunspot region. 
}
\end{figure}

To examine the change of magnetic fluxes
before and after the flare occurrence,
we have plotted the time variation of magnetic fluxes of positive and negative
polarities in Figure  2, in which  the asterisked  curves show 
the  fluxes in a $\delta$ sunspot region. 
As seen  in the  figure,  magnetic  fluxes of 
the $\delta$  sunspot region for both 
polarities decreased with time.
It is also noted that there were several emerging fluxes (P1, P2,
N1, N2 in Figure 1) outside
the $\delta$ sunspot region, which were
accompanied by eruptions of $H\alpha$ arch filaments (Canfield and Reardon, 1998). The new emerging positive fluxes such as P1 and P2 
account for total flux increase for positive polarities.
These facts imply that 
the X-class flare should be associated with both 
 cancellation of the $\delta$-type magnetic fields and new emerging fluxes.

\begin{figure}
\psfig{figure=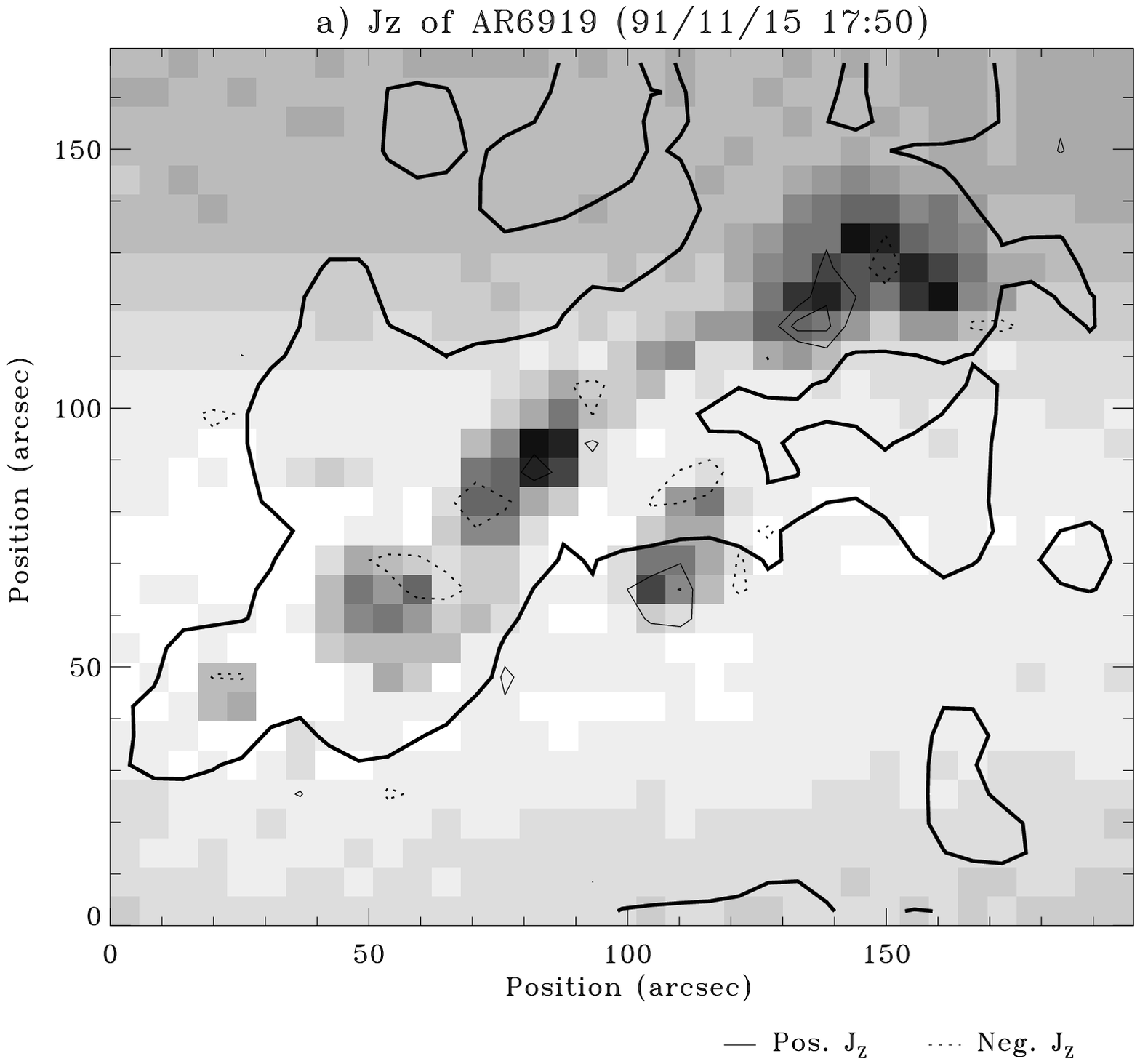,height=5.8cm,width=8cm}
\psfig{figure=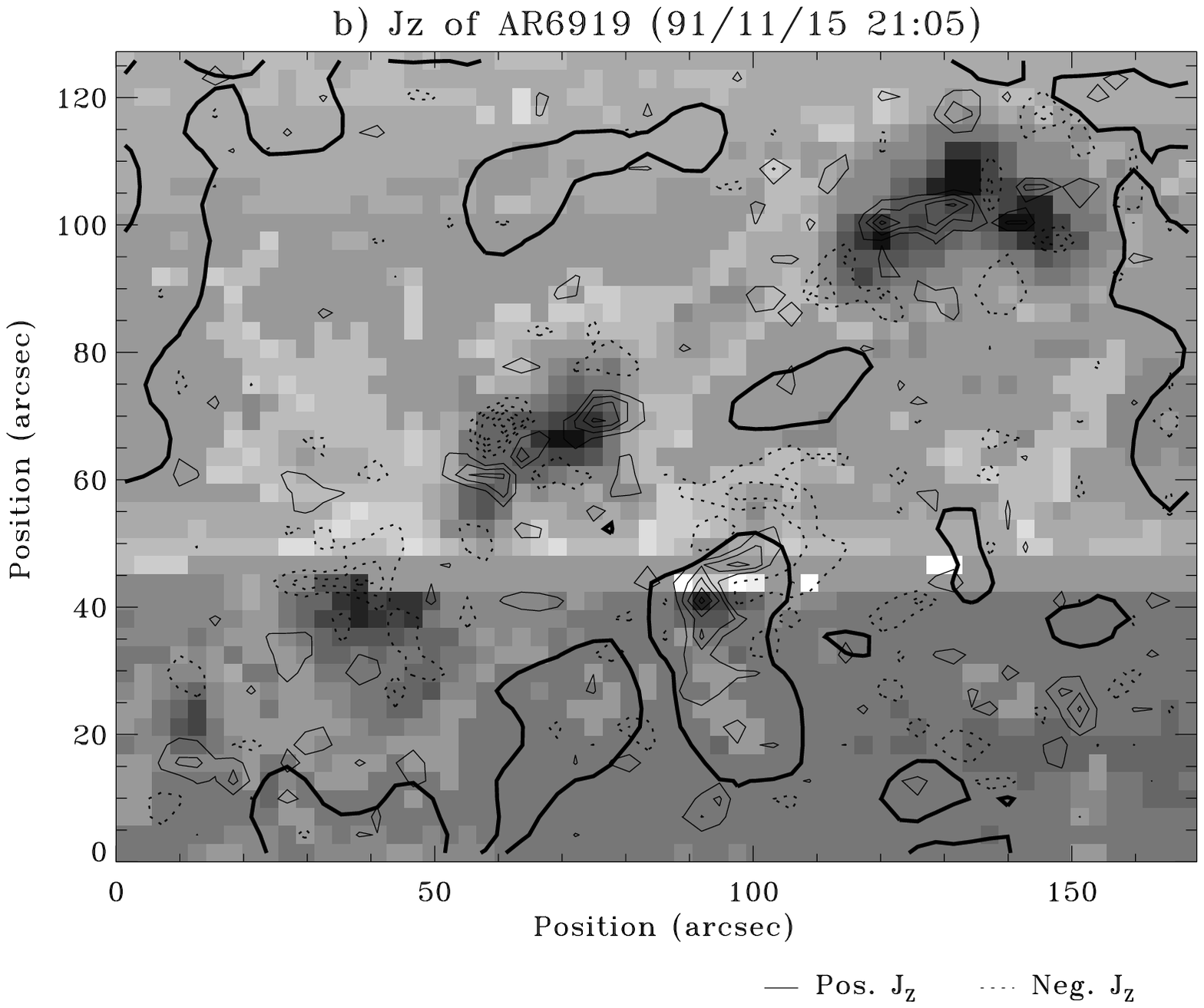,height=5.8cm,width=8cm}
\psfig{figure=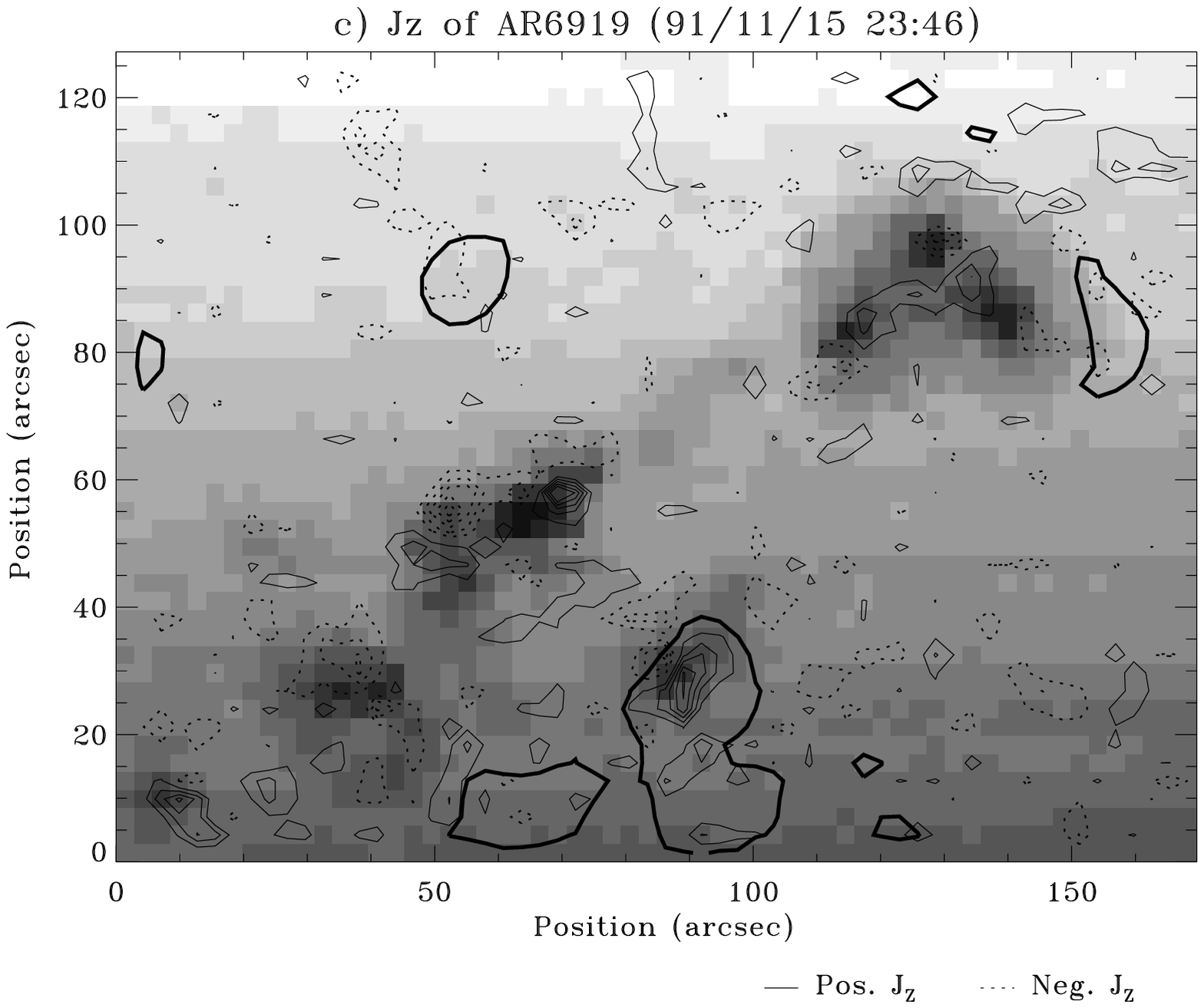,height=5.8cm,width=8cm}
\caption{ Vertical current densities  
of AR 6919  superposed on white light images.  
The contour levels correspond to
3, 6, 9, 12 and 15  $\rm mA/m^2$,
respectively.
Thick solid lines represent the
inversion lines inferred from the longitudinal magnetic field data.
}
\end{figure}

We show the vertical current density and  inversion lines inferred
from the longitudinal fields in Figure 3.
As seen in Figures 3-b) and 3-c),  the strongest vertical current
density kernel is  located near the inversion line of
the $\delta$ sunspot.
As seen in the figures, the vertical current density much 
increased just before the onset of the X-class flare.

\begin{figure}
\psfig{figure=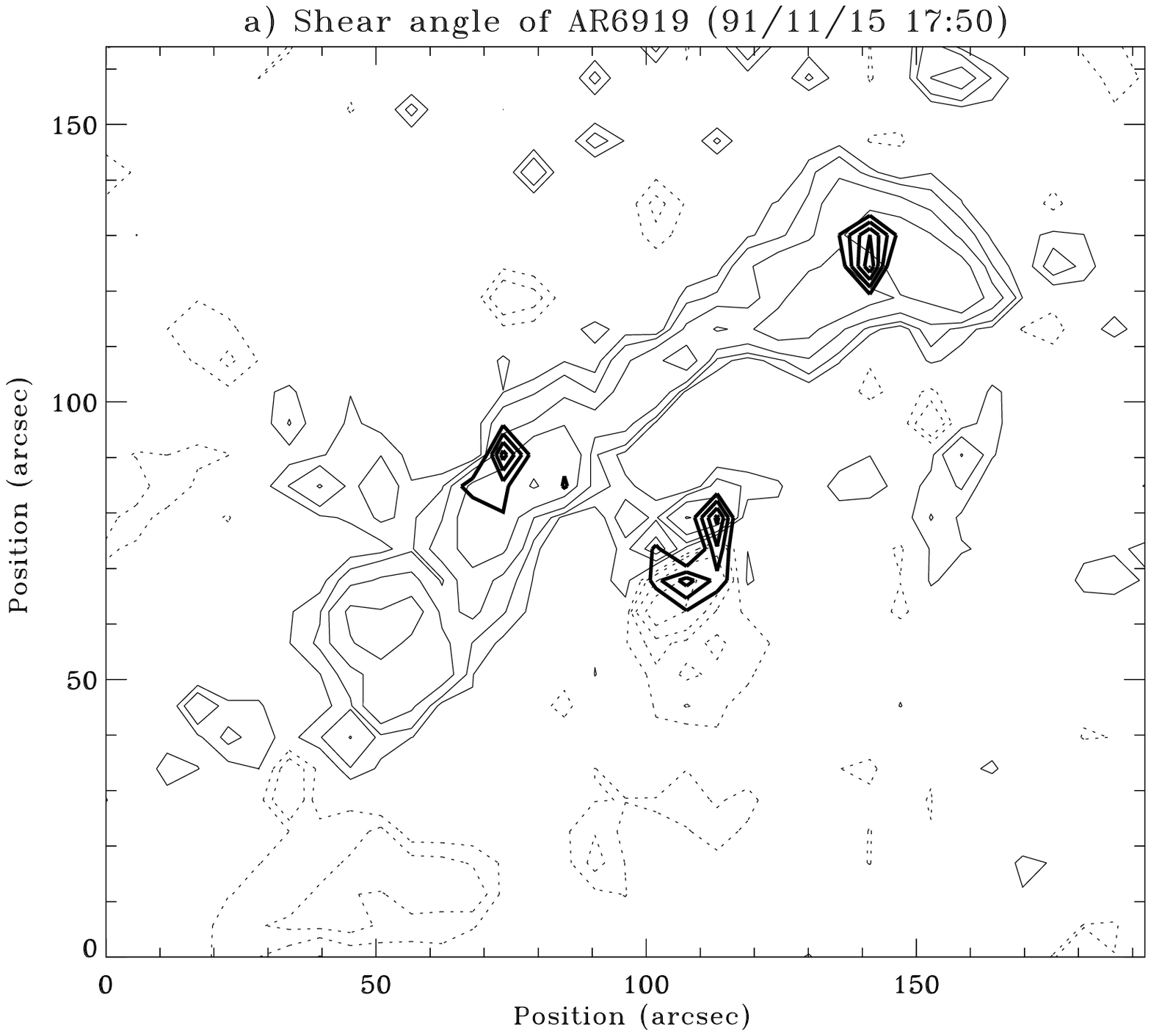,height=6cm,width=8.5cm}
\psfig{figure=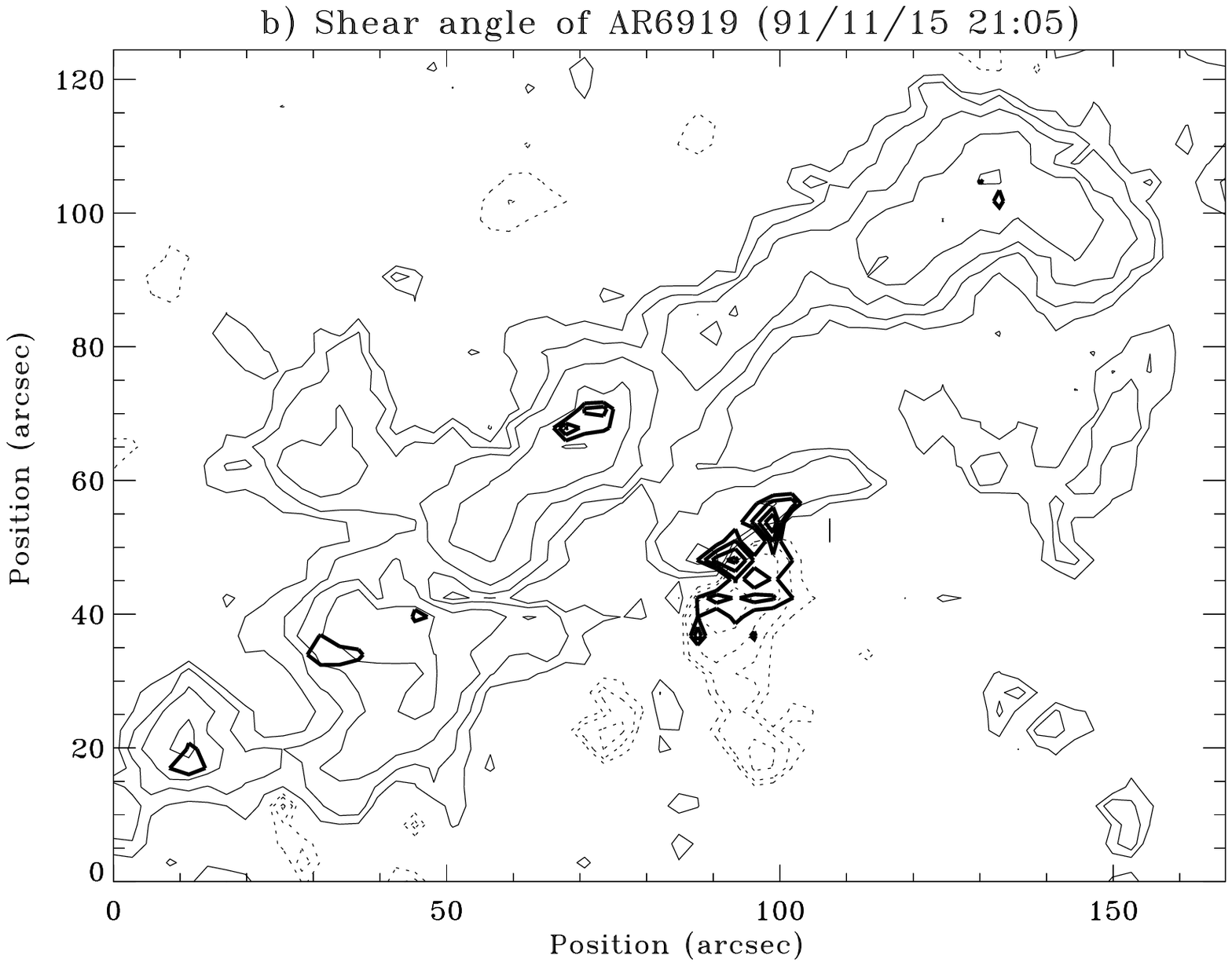,height=6cm,width=8.5cm}
\psfig{figure=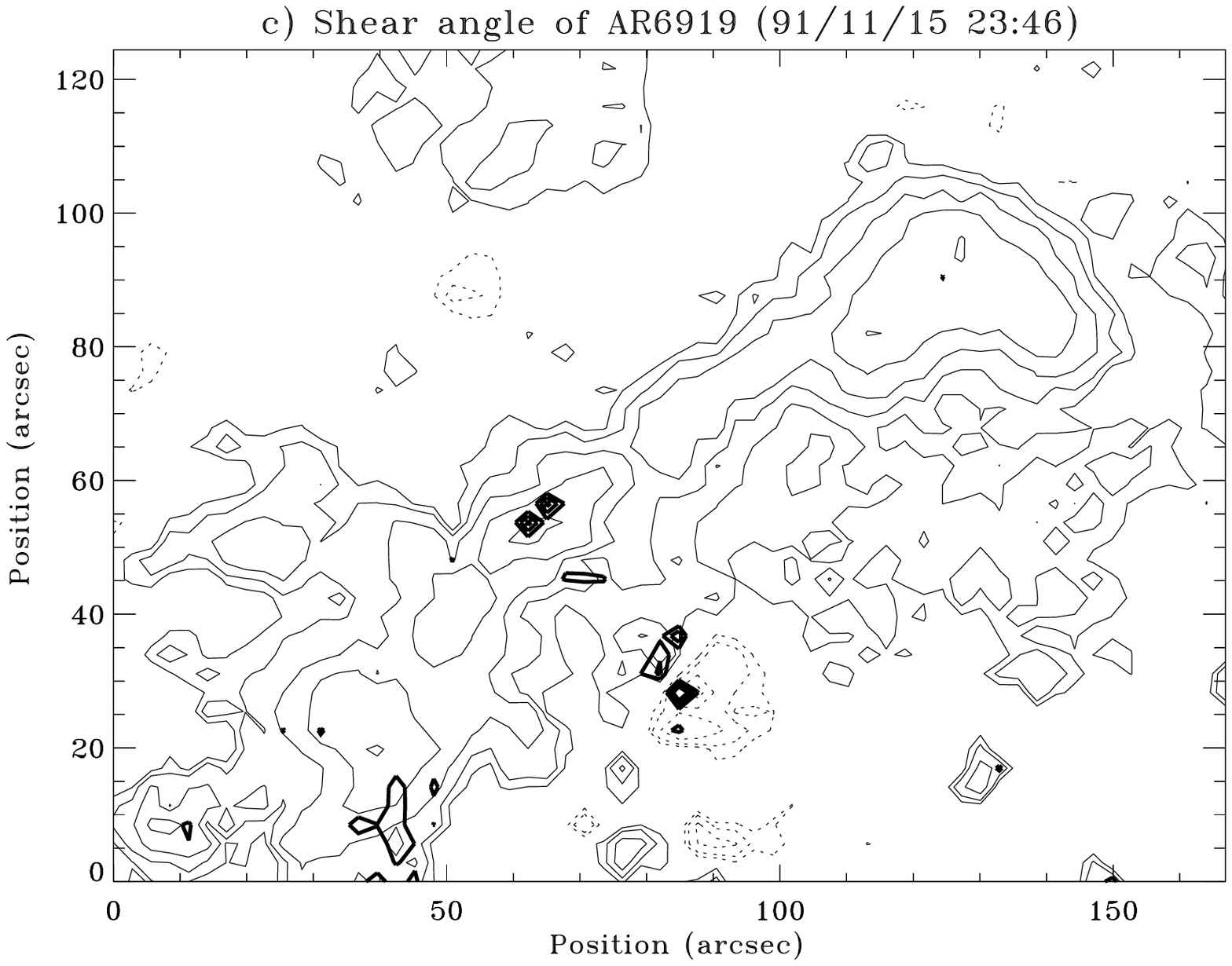,height=6cm,width=8.5cm}

\caption{Contours of angular shear multiplied by horizontal field
strength obtained from three vector magnetograms of AR 6919 on 
Nov. 15, 1991 drawn in thick solid lines are superposed on the vertical
magnetograms. 
(thick solid lines) 
The contour levels are 30000, 40000, 50000 and 60000 G deg,
respectively.
In all the figures, the thin solid lines stand for positive
polarities and the thin dotted lines for 
negative polarities. 
The contour levels in the
magnetograms correspond to  100, 200, 400, 800 and 1600 G,
respectively.}
\end{figure}

\begin{figure}
\psfig{figure=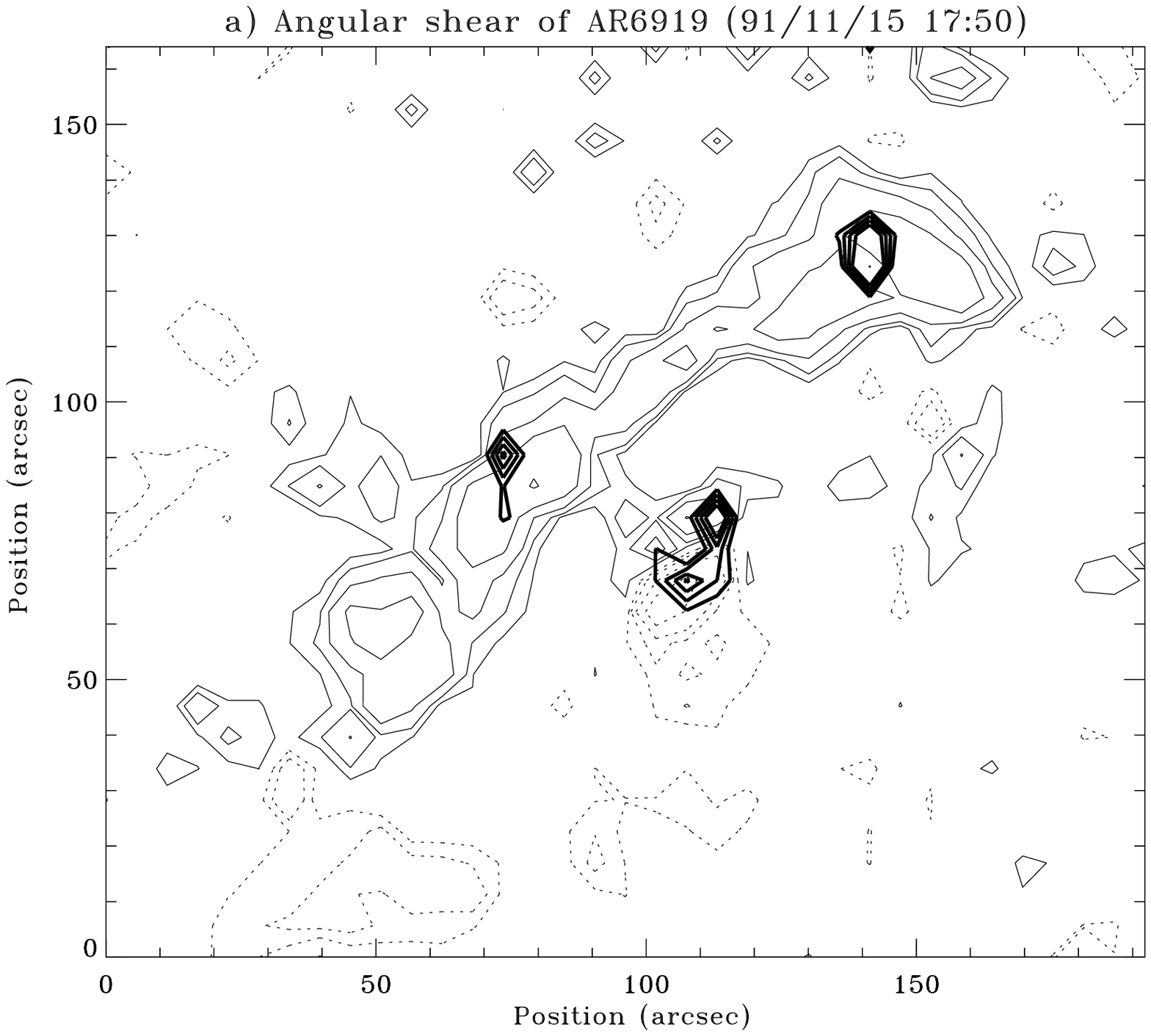,height=6cm,width=8.5cm}
\psfig{figure=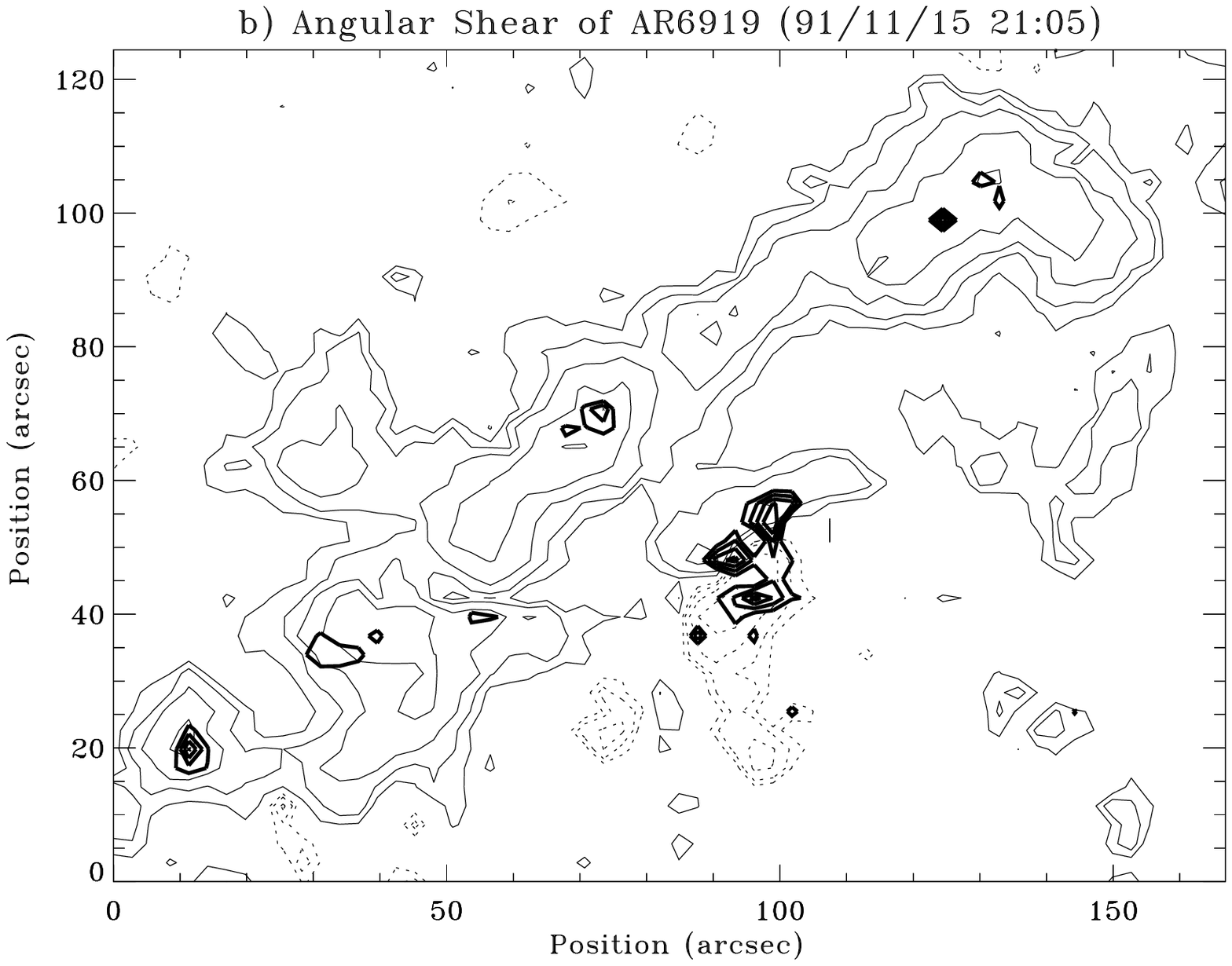,height=6cm,width=8.5cm}
\psfig{figure=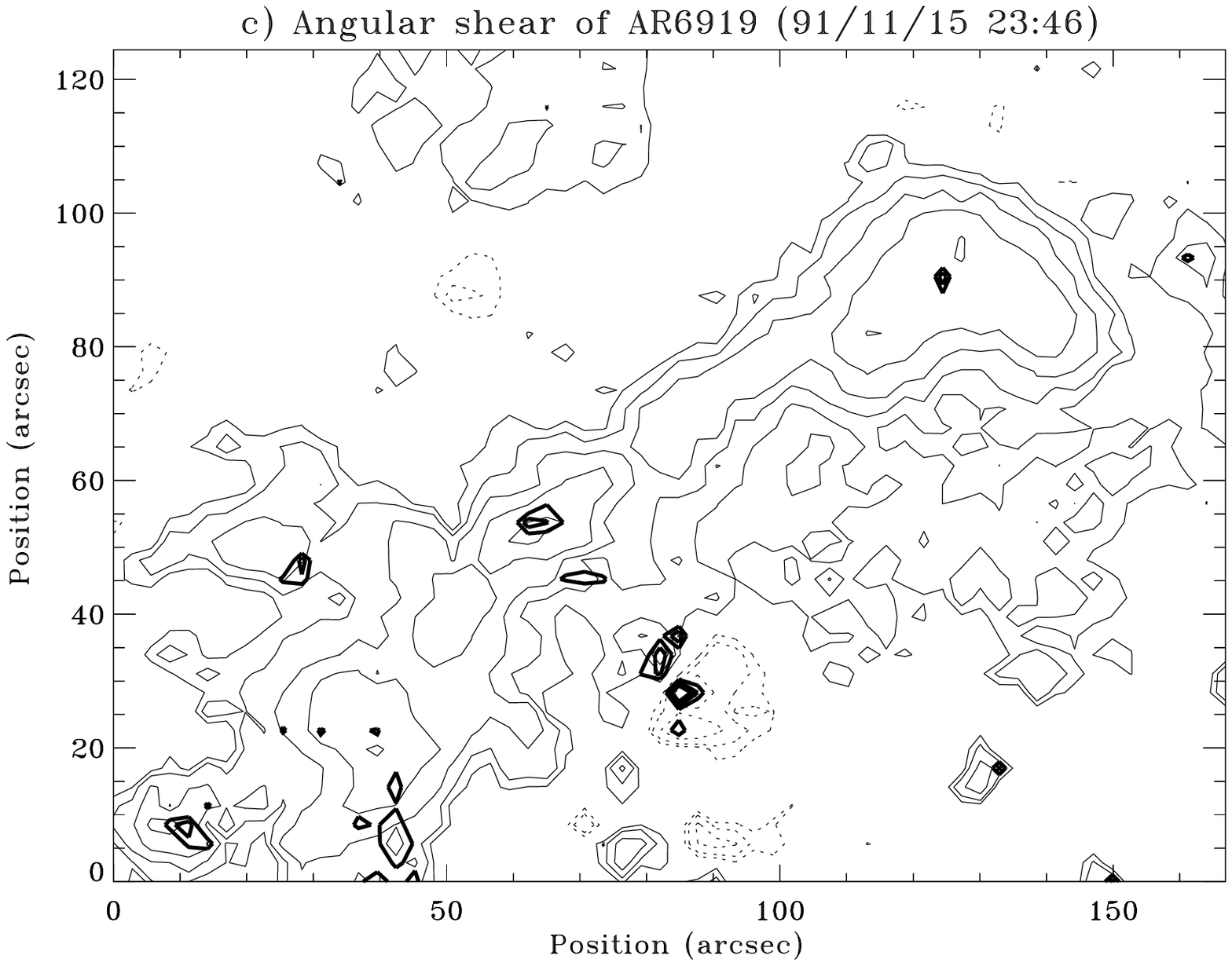,height=6cm,width=8.5cm}
\caption{Contours of shear angle multiplied by field strength
drawn in thick solid lines are superposed on the vertical 
magnetograms. 
The contour levels are 30000, 40000, 50000 and 60000 G deg,
respectively. 
 The magnetograms are the same as in Figure 4.
}
\end{figure}

\begin{figure}[t]
\psfig{figure=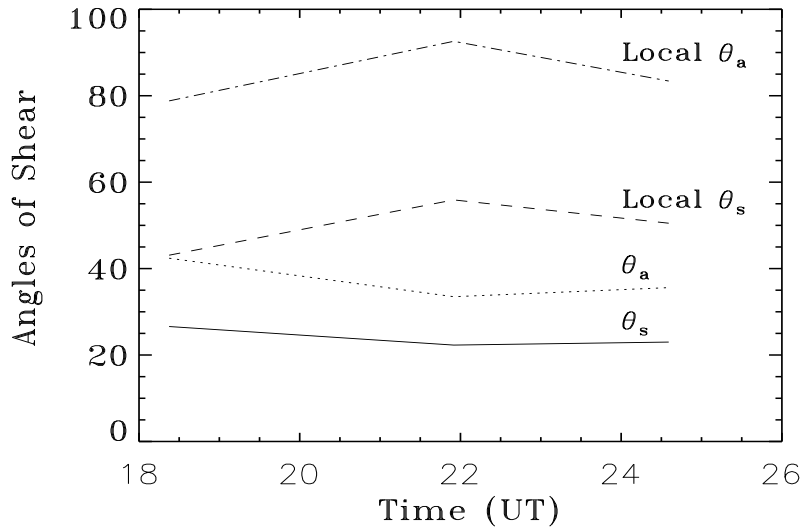,height=8cm,width=12cm}
\caption{Time variation of weighted shear angle and
angular shear before and
after an X-class flare in AR 6919. In the figure Local 
$\theta_a$ and Local $\theta_s$ are weighted shear angles for the $\delta$ sunspot region.
}
\end{figure}

\begin{table}
\caption{Weighted mean
magnetic shear angle, angular 
shear and sum of magnetic free energy density for AR 6919 
for three different times. All parameters are calculated 
for pixels with $B_z > 100 G$. The asterisked data are obtained 
adopting the potential field method for $180\deg$ ambiguity 
resolution.} 
\begin{tabular}{ccccccccc} \hline
\hline
Data  &  $\overline \theta_s$ &  local $\overline \theta_s $  & 
$\overline \theta_a$  & local  $\overline \theta_a$  & $\overline 
 \rho_f [{\rm erg/cm^3}]$  & 
local $\overline \rho_f$
& $\sum \rho_f [{\rm erg/cm}]$ & local $\sum \rho_f $  \\
\hline
a) &  26.6 & 43.1 & 42.4 & 78.8 & 5.0E3 & 1.6E4 &1.6E23 & 3.7E22\\
b) &  22.3 & 55.9&33.5&92.6&3.6E3&1.6E4&1.1E23&2.6E22\\
c) &  23.0 & 50.5& 35.6&83.4&3.0E3&7.9E3&1.3E23&1.3E22\\
a)* &  23.3 & 30.1&32.4&48.1&4.1E3&9.3E3&1.3E23&2.2E22\\
b)* &  20.2 & 36.3&28.3&51.2&2.9E3&8.0E3&9.2E22&1.3E22\\
c)* &  21.3 & 36.8&30.8&56.2&2.6E3&4.7E3&1.1E23&7.8E21\\
\hline
\end{tabular}
\end{table}

\begin{figure}
\psfig{figure=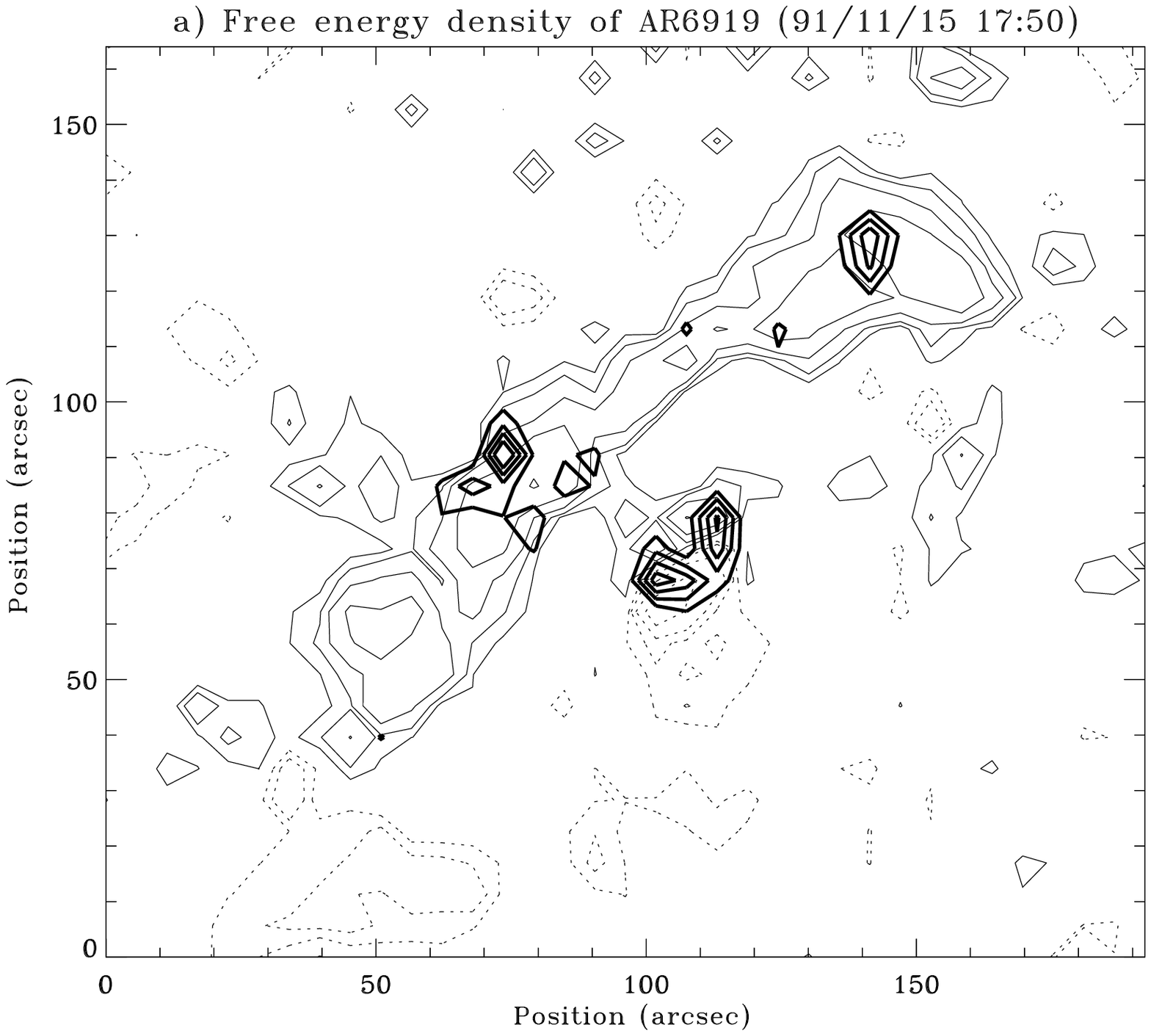,height=6cm,width=8.5cm}
\psfig{figure=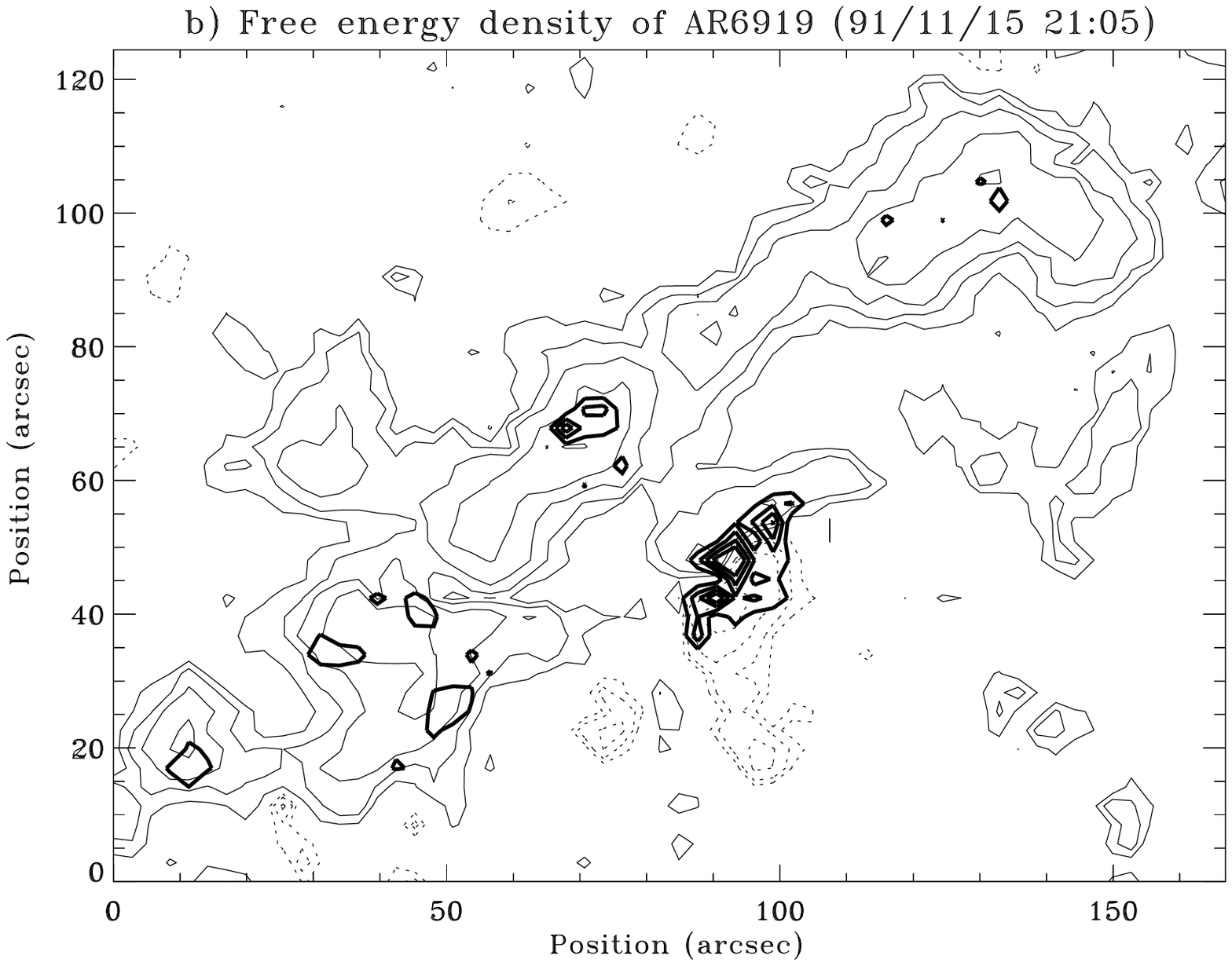,height=6cm,width=8.5cm}
\psfig{figure=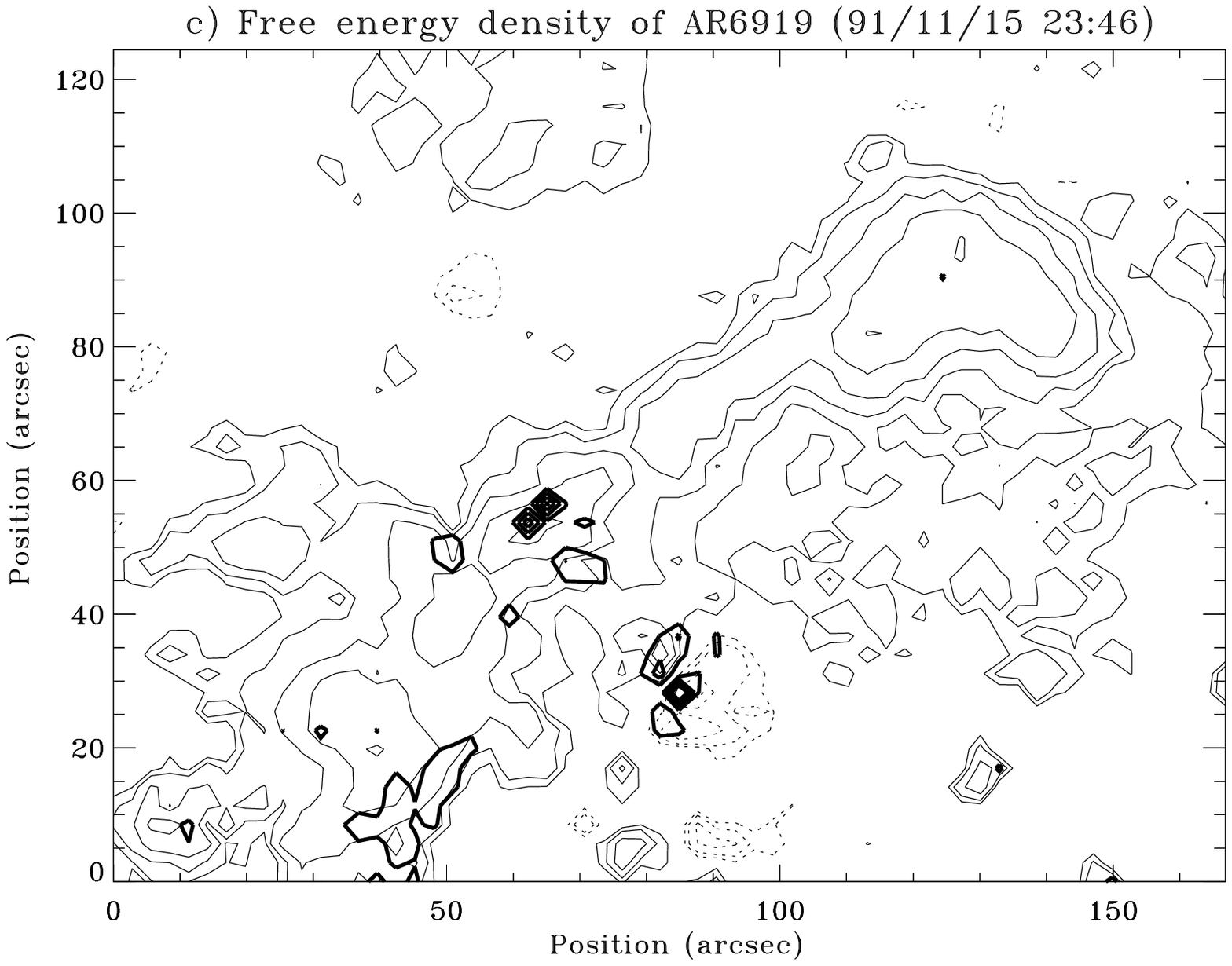,height=6cm,width=8.5cm}
\caption{Contours of free energy density obtained from  
three vector magnetograms of AR 6919 drawn in thick solid lines are 
superposed on the vertical magnetograms. 
The contour levels are $1\times 10^4$, $2\times 10^4$, $3\times 10^4$ and 
$4\times 10^4\, \rm erg/cm^3$,
respectively.
The magnetograms are the same as in Figure 4.
}
\end{figure}

\begin{figure}[t]
\psfig{figure=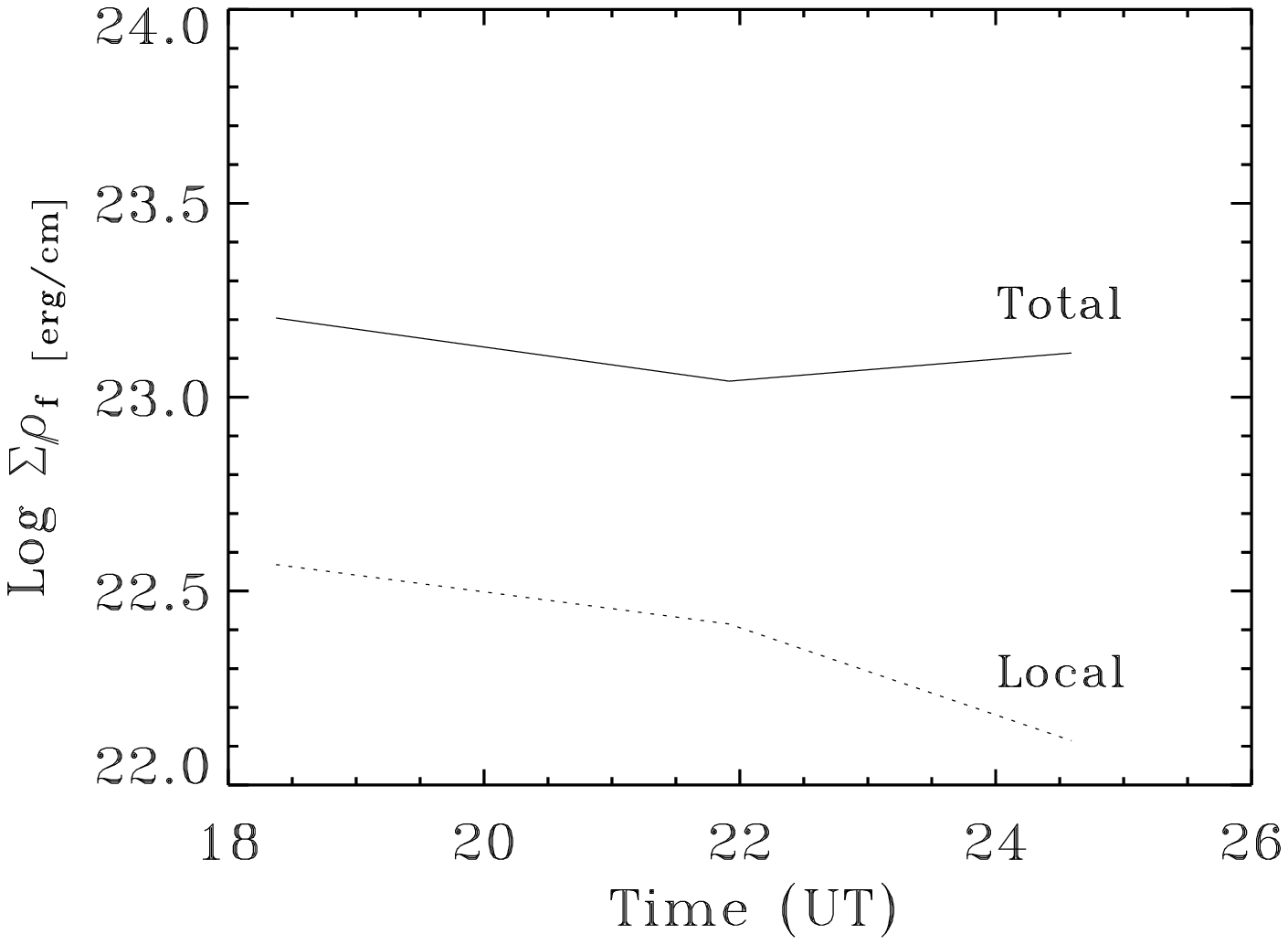,height=8cm,width=12cm}
\caption{Time variation of the planar sum of free energy density
before and
after an X-class flare of AR 6919. In the figure Local corresponds to a $\delta$ sunspot region.
}
\end{figure}

Figures 4 and 5 show angular shear  weighted with transverse field strength and shear 
angle weighted   with total   field strength   derived from  vector  magnetograms  and 
computed potential fields (Sakurai, 1992). 
As seen in the figures,
high values of shear angles are located near the $\delta$
spot region. The changes of two
mean shear angles with time are given in Figure 6, which shows that shear angles for
the $\delta$ spot area (denoted by Local in the figure) increased before the X-class flare and then decreased 
after it, but those for
the whole active region  decreased  before the flare. 
It is to be noted that a similar pattern is also found in the
evolution of a measure of magnetic field discontinuity, MAD
(see Figure 7 in  Moon et el. 1999a). 
The derived  mean shear  angles are  listed in  Table II,  in which for  comparison we  
also tabulate  the corresponding values obtained with the potential field method for 
$180\deg$ ambiguity resolution. 
It is interesting  that
the mean  shear  angles for   the $\delta$ spot  area (denoted by local in Table II)
obtained with the  potential  
method continuously 
increased during the flaring activity unlike those obtained with our method.
This eloquently demonstrates  that totally  different conclusions  can be  drawn on  the 
shear angle  evolution  throughout a   flare, critically  depending on   the $180\deg$ 
ambiguity resolution method employed in data reduction. 

The magnetic free energy densities
are shown in Figure 7. 
The free energy density  near the $\delta$ spot region
largely decreased before the flare.
We also list mean energy densities and total energy densities in Table II and
plot their variations with time  in Figure 8. These data indicate that
magnetic free energy, at least in the neighborhood of the $\delta$ spot, 
was indeed released during the flaring process.

It is notably interesting and not easy to understand 
that in Figures 6 and 8, the curvatures
of curves for the whole active region 
are of opposite sign to those for the local $\delta$ 
spot region. This may create some skepticism about the conventional picture of the flare 
mechanism. 
In Figure 6, where shear angle variations 
are displayed, the curves (Local $\theta_a$ and Local $\theta_s$) for the $\delta$ spot region  seem to be consistent with the 
conventional flare picture, according to which the magnetic shear increases to build up
free energy until the flare onset and then  
decreases as free energy being released. 
The free energy curve for the same 
$\delta$ spot region in Figure 8, however, shows 
that the energy decrease started even before the flare onset
although a more drastic
energy release followed the onset. 
This indicates that relaxation of the twisted magnetic
field mildly starts even hours before the visible flare onset. 
The decrease of mean shear 
angle for the whole active region before the flare supports this argument. Why, then, 
does the local shear increase before the flare? This may be interpreted as development 
of a current sheet of a macroscopic size in the possible reconnection region near the
$\delta$ spot. Magnetic fields tend to take a lowest possible energy state under given 
constraints. The lowest energy state  is expected to have a  smooth spatial variation of 
magnetic field. However, a smooth state may not exist if the field has undergone 
too much twist.   Then, a current  sheet must develop in some 
regions  while the field 
takes a smooth configuration in other regions. The process of current sheet 
development may be either a pure ideal MHD process as described above or a process 
involving a lot of small scale magnetic reconnections. The latter process  
can be analogized 
with an avalanche of a sand pile initiated by a slide of a few sand grains. 
This scenario was proposed and investigated by Lu {\it et al.} (1991, 1993). 
The increase of  the planar sum  of free energy  density over the  whole active region 
shown in Figure 8 still 
requires more investigation. However, this can be also explained within the conventional 
picture of solar flares. When a magnetic field is twisted, the whole field tends to expand 
to occupy more volume per flux. 
The magnetic energy is thus more concentrated near the surface 
in a potential field than in a twisted field. The increase of the free energy after the 
flare onset might imply that the coronal field shrinks down while only part of 
the total free energy is released by the flaring process. However, our speculations based 
on only one flare case study definitely need to be examined by investigation of 
more cases.


\section{Summary and Conclusion}
In this study,  
we have studied the magnetic nonpotentiality 
of AR 6919 associated with an X-class flare which occurred on November
15, 1991, using MSO magnetograms. 
The magnetogram data were obtained before and after the X-class flare
using the Haleakala Stokes Polarimeter. A nonlinear least square
method was adopted to derive the magnetic field components from the
observed Stokes profiles and a multi-step ambiguity solution
method  was employed to resolve the
$180\deg$ ambiguity. From the $180\deg$ 
ambiguity-resolved vector magnetograms,
we have derived a set of physical quantities characterizing 
the field configuration.
The results emerging from this study can be summarized as follows:

1) There was flux decreases for  both polarities  in a $\delta$ sunspot 
pair as well as  flux increases outside it, which implies that the  energy release of  the X-class flare  should  be associated  
with flux cancellation and emergence. 

2) It was  found that the  vertical current  near the $\delta$  sunspot much increased 
before the flare.

3) We also found  that  magnetic shear  near the  $\delta$ sunspot  increased before 
the flare and then decreased after it, while magnetic shear in the whole active region
decreased before the flare.

4) The sum of magnetic energy density  decreased before the flare, indicating that
magnetic free energy was released by
the flaring event.

However, we also found different evolutionary tendencies of nonpotentiality parameters
for the whole active region and for the local $\delta$ spot region. These differences
must be examined through further studies of many other flare-bearing active regions.  
If some of them could be confirmed, these would serve as important 
clues to understand flaring mechanisms.

\acknowledgements
 We wish to thank Dr. Metcalf 
 for allowing us to use some of his numerical routines
 for analyzing vector magnetograms and Dr. Labonte for helpful comments.
The data from the Mees Solar Observatory, University of Hawaii 
are produced with the support of NASA grant NAG 5-4941 and
NASA contract NAS8-40801.
This work has been supported in part by
the Basic Research Fund (99-1-500-00 and 99-1-500-21) of 
Korea 
Astronomy Observatory and in part by the Korea-US Cooperative Science Program under KOSEF(995-0200-004-2).



\begin{thebibliography}{}

\bibitem[Ambastha {\it et al.} 1993]{amb93}
Ambastha, A., Hagyard, M. J., and  West, E. A.: 1993, {\it Solar Phys.} {\bf 148}, 277.

\bibitem[Canfield {\it et al.} 1992]{can92}
Canfield,   R. C., Hudson, H. S., Leka, K. D., Mickey, D. L.,
Metcalf, T. R., W\"ulser,
 J-P., Acton, L. W.,
Strong, K. T., Kosugi, T., Sakao, T., 
Tsuneta, S.,
Culhane, J. L., Phillips, A.,
 and Fludra, A.: 1992,         
{\it Pub. Astron. Soc. Japan} {\bf 44}, L111.

\bibitem[Canfield {\it et al.} 1993]{can93}
Canfield,   R. C.,
La Beaujardiere, J.-F., Han, Y., Leka, K. D., 
McClymont, A. N.,
Metcalf, T. R., Mickey, D. L., W\"ulser, J.-P., and Lites, B. W.:
1993,       
{\it Astrophys. J.} {\bf 411}, 362.

\bibitem[Canfield and Reardon 1998]{can98}
Canfield,   R. C. and Reardon, K. P.:
1998,       
{\it Solar Phys.} {\bf 182}, 145.


\bibitem[Chen {\it et al.} 1994]{che94}
Chen, J., Wang, H., Zirin, H, and 
Ai, G.: 1994, {\it Solar Phys.} {\bf 154}, 261.



\bibitem[Gary {\it et al.} 1987]{gar87}
Gary, G. A., Moore, R. L., Hagyard, M. J., and
Haisch, B. M.: 1987, {\it Astrophys. J.}
{\bf 314}, 782.



\bibitem[Gary {\it et al.} 1991]{gar91}
Gary, G. A., Hagyard, M. J., and West, E. A.: 1991, in L. November (ed.),
{\it Solar Polarimetry}, Proceeding of the Workshop of Solar
Polarimetry, p. 65.



\bibitem[Hagyard {\it et al.} 1981]{hag81}
Hagyard, M. J., Low, B. C., and Tandberg-Hanssen, E.: 1981, {\it Solar Phys.} {\bf 73}, 257.





\bibitem[Hagyard {\it et al.} 1982]{hag82}
Hagyard, M. J., Cummings, N. P., West E. A., and Smith, Jr., J. B.:
1982, {\it Solar Phys.} {\bf 80}, 33.





\bibitem[Hagyard {\it et al.} 1984]{hag84}
Hagyard, M. J.,  Smith, Jr., J. B., Teuber, D., and West, E. A.: 1984, 
{\it Solar Phys.} {\bf 91}, 115.



\bibitem[Hagyard {\it et al.} 1988]{hag88}
Hagyard, M. J.,  Gary, G. A., and West, E. A.: 1988, {\it
The SAMEX Vector Magnetograph},
NASA Technical Memorandum 4048



\bibitem[Hagyard {\it et al.} 1990]{hag90}
Hagyard, M. J., Ventkatarishnan, P., 
and Smith, Jr., J. B.: 1990, {\it Astrophys. J. Suppl.} {\bf 73}, 159.



\bibitem[Hagyard {\it et al.} 1993]{hag93}
Hagyard, M. J., West, E. A., and Smith,  J. E.: 1993, {\it Solar Phys.} {\bf 144},141.



\bibitem[Hagyard and Kineke 1995]{hag95}
Hagyard, M. J. and Kineke, J. I.: 1995, {\it Solar Phys.} {\bf 158}, 11.

\bibitem[Hudson {\it et al.} 1992]{hud92}
Hudson, H. S., Acton, L. W., Hirayama, T., and Uchida, Y.: 1992,
 {\it Pub. Astron. Soc. Japan} {\bf 44}, 77.


\bibitem[Jefferies and Mickey 1991]{jef91}
Jefferies, J. T. and Mickey, D. L.: 1991, {\it Astrophys. J.} {\bf 372}, 694.



\bibitem[Kim 1997]{kim97}
Kim, K. S.: 1997, {\it Pub. Korean Astron. Soc.} {\bf 12}, 1.


\bibitem[Lu {\it et al.} 1991]{luet91}
Lu, E. T. and Hamilton, R. J.: 1991, {\it Astrophys. J.} {\bf 380}, L89.

\bibitem[Lu {\it et al.} 1993]{luet93}
Lu, E. T., Hamilton, R. J., McTiernan, J. M., and Bromund, K. R.:
1993, {\it Astrophys. J.} {\bf 412}, 841.



\bibitem[L\"u {\it et al.} 1993]{lu93}
L\"u, Y., Wang, J., and Wang, H.: 1993, {\it Solar Phys.} {\bf 148}, 119.



\bibitem[McClymont, Jiao, and Micki\'c 1997]{mcc97}
McClymont, A. N. Jiao, L., and Micki\'c, Z.: 1997, {\it Solar Phys.}
{\bf 174}, 191.

\bibitem[Metcalf {\it et al.} 1995]{metcalf95}
Metcalf, T. R., Jiao, L., McClymont, A. N., Canfield, R. C., and
Uitenbroek, H.: 1995, {\it Astrophys. J.} {\bf 439}, 474.


\bibitem[Mickey 1985]{mic85}
Mickey, D. L.:   1985,   
{\it Solar Phys.}
{\it 97}, 223.



\bibitem[Moon {\it et al.} 1999a]{moo99a}
Moon, Y.-J., Yun, H. S., Lee, S. W., Kim, J.-H. Choe, G. S., Park, Y .D., 
Ai, G., Zhang, H., and Fang, C.: 1999a, {\it Solar Phys.} {\bf 184}, 323.

\bibitem[Moon {\it et al.} 1999b]{moo99b}
Moon, Y.-J., Park, Y. D., and Yun, H. S.: 
1999b, {\it J. Korean Astron. Soc.} 
{\it 32}, 63.

\bibitem[Moon {\it et al.} 1999c]{moo99c}
Moon, Y.-J., Yun, H. S.,  Choe, G. S., Park, Y .D., 
Mickey, D. L.: 1999c, submitted to Solar Physics.



\bibitem[Ronan {\it et al.} 1987]{ron87}
Ronan, R. S., Mickey, D. L., and Orral, F. Q.:  1987,
{\it Solar Phys.} {\bf 113}, 353.

\bibitem[Sakao 1992]{saka92}
Sakao, T., Kosugi, T., Masuda, S., Inda, M, Makishima, K.,
Canfield, R. C., Hudson, H. S., Metcalf, T. R., W\"ulser, J.-P.,
Acton, L. W., Ogawara, Y.:
 1992, {\it Pub. Astron. Soc. Japan} {\bf 44}, L83.

\bibitem[Sakurai  1992]{sak92}
Sakurai, T.: 1992, in his numerical routines

\bibitem[Sakurai {\it et al.} 1995]{sak95}
Sakurai, T., Ichimoto, K,. Nishino, Y., Shinoda, K., Noguchi, M., Hiei, E.,
Li, T., He, F., Mao, W., Lu, H., Ai, G., Zhao, Z., Kawakami, S., and
Chae, J.:  1995, {\it Pub. Astron. Soc. Japan} {\bf 47}, 81.


\bibitem[Skumanich and Lites 1987]{sku87}
Skumanich, A. and Lites, B. W.: 1987, {\it Astrophys. J.} {\bf 322}, 473.



\bibitem[Wang 1992]{wan92}
Wang, H.: 1992, {\it Solar Phys.}  {\bf 140}, 85.



\bibitem[Wang and Tang 1993]{wan93}
Wang, H. and Tang, F.: 1993, {\it Astrophys. J. } {\bf 407}, L89.

 

\bibitem[Wang {\it et al.} 1994]{wan94}
Wang, H., Ewell, W., Zirin, H., and Ai, G.: 1994, {\it Astrophys. J.} {\bf 424}, 436.



\bibitem[Wang 1997]{wan97}
Wang, H.: 1997, {\it Solar Phys.} {\bf 174}, 163.





\bibitem[Wang {\it et al.} 1996]{wan96}
Wang,  J., Shi,  Z., Wang, H., and Lu, Y.:
1996,  {\it Astrophys. J.} {\bf 456},  861.

\bibitem[W\"ulser {\it et al.} 1994]{wul94}
W\"ulser, J.-P., Canfield, R. C., Actor, L. W., 
Culhane, J. L., Philips, A., Fludra, A., Sakao, T., Masuda, S.,
Kosugi, T., and Tsuneta, S.: 1994, {\it Astrophys. J.} {\bf 424}, 459.


\bibitem[Zirin 1995]{zir95}
Zirin, H.: 1995, {\it Solar Phys.} {\it 159}, 203.

\end{thebibliography}
\end{document}